\documentclass[epj]{svjour}
\usepackage{amssymb,psfig}
\begin{document}
\title{ 
{\normalsize \rm Eur. Phys. J. B {\bf 22}, 291--300 (2001)}\\
\\
\\
\\
\\
\\
{Spin-wave theory for finite classical magnets \\
and superparamagnetic relation}
}
\author{H. Kachkachi$^*$ \and D. A. Garanin$^\dagger$ 
}                     
\institute{Laboratoire de Magn\'etisme et d'Optique, Universit\'e de Versailles St. Quentin, 45 av. des \'Etats-Unis, 78035 Versailles, France}
\date{Received: 1 December 2000 }
%
\abstract{
Analytical calculations based on finite-size spin-wave theory and Monte Carlo (MC) simulations are performed to investigate the validity of the well-known relation $m(H,T)=M(H,T)B_{D}[M(H,T){\cal N}H/T]$  between the induced magnetization $m$ of the magnetic particle and its intrinsic magnetization $M$ for the Ising and isotropic classical models [$B_{D}(x)$ is the Langevin function, $D$ is the number of spin components, ${\cal N}$ is the number of atoms in the particle]. It follows from general arguments and from our analytical results for the Heisenberg model at $T\ll T_{c}$ that this relation is {\em not exact} for any finite $D$ and nonzero temperature. Nevertheless, corrections to this formula remain very small practically in the whole range $T<T_{c}$ if ${\cal N}\gg 1$, as confirmed by our Monte Carlo calculations. At $T\lesssim T_{c}/4$ there is a good agreement between the MC and finite-size spin-wave calculations for the field dependence of $m$ and $M$ for the Heisenberg model with free boundary conditions.
} 
\authorrunning{H. Kachkachi and D. A.Garanin}
\titlerunning{Spin-wave theory for finite classical magnets and superparamagnetic relation}
\maketitle

\section{Introduction}

\label{SecIntroduction}

For magnetic particles of a finite size one can generally define two
magnetizations, $m$ and $M,$ the relation between which is frequently
written in the form 
\begin{equation}
m=MB_{D}(Mx),\qquad x\equiv {\cal N}H/T,  \label{spmrelation}
\end{equation}
where $B_{D}(x)$ is the Langevin function [$B_{D}(x)=B_{3}(x)=\coth x-1/x$
for the isotropic Heisenberg model and $B_{D}(x)=B_{1}(x)=\tanh x$ for the
Ising model] and ${\cal N}$ is the number of magnetic atoms in the system.
Here $m$ is the magnetization induced by the magnetic field and
microscopically defined as the thermodynamic average of the vector 
\begin{equation}
{\bf M}=\frac{1}{{\cal N}}\sum_{i}{\bf s}_{i},  \label{MvecDef}
\end{equation}
i.e.,

\begin{equation}
{\bf m}=\langle {\bf M}\rangle .  \label{mdef}
\end{equation}
For classical systems discussed throughout this paper, ${\bf s}_{i}$ can be
considered, up to a factor, as spin vectors of unit length, $|{\bf s}%
_{i}|=1. $ The magnetization $M$ in Eq. \negthinspace (\ref{spmrelation})
can be interpreted as the intrinsic magnetization of the particle which is
defined through the correlation function of the magnetic moments,

\begin{equation}
M=\sqrt{\left\langle {\bf M}^{2}\right\rangle }  \label{Mdef}
\end{equation}

If the temperature is low, all spins in the particle are bound together by
the exchange interaction and ${\bf M}$ behaves as a rigid ``giant spin'', 
\mbox{$\vert$}%
${\bf M|\cong }M{\bf \cong }1,$which shows a {\it superparamagnetic}
behavior. If a magnetic field ${\bf H}$ is applied, ${\bf M}$ exibits an
average in the direction of ${\bf H,}$ which leads to a nonzero value of the
induced magnetization ${\bf m}$ given, obviously, by Eq. \negthinspace (\ref
{spmrelation}). The question of principal interest is, however, the field
dependence of $M$ at nonzero temperatures, which can be responsible for
deviations from the simple superparamagnetic behaviour of Eq. \negthinspace (%
\ref{spmrelation}).

Early Monte Carlo (MC) simulations by Wildpaner \cite{wil74} for the
classical Heisenberg model, where both magnetizations were determined
independently as functions of field at different temperatures, confirmed Eq.
\negthinspace (\ref{spmrelation}) within numerical errors. However, from the
theoretical point of view this relation with $M=M(H,T)$ is unexpected.

On the theoretical side, Eq. \negthinspace (\ref{spmrelation}) was obtained
in Ref. \negthinspace \cite{hasleu90} for a classical model and in Ref.
\negthinspace \cite{hasnie93} for a quantum model but without the field
dependence of $M$. Earlier, Fisher and Privman \cite{fispri85} considered the
spin-wave contribution to Eq. \negthinspace (\ref{spmrelation}) but, again,
the field dependence of $M$ was not studied explicitly.

Experimentally, the field dependence of $M$ and, in particular, the
nonsaturation of the magnetization in the region $x\gtrsim 1$ have been
observed in nanoparticles by different groups \cite
{dorfiotro97,chesorklahad95,resetal98}. Usually this dependence is close to
linear and is used to extract the value of $M$ at zero field by
extrapolation to $H=0$. For the isotropic Heisenberg model, the field
dependence of $M$ in the range $x\gtrsim 1$ is due to suppression of the
fluctuations of individual spins, i.e., of spin waves, and this dependence
disappears for $T\rightarrow 0$. The dependence $M(H)$ is much stronger and
persists at zero temperatures if the spins in the particle are not perfectly
collinear due to surface effects\cite{resetal98}.

In our recent paper \cite{kacgar01} (see also Ref.\negthinspace\ \cite
{kacnogtrogar00}) we have shown that this relation becomes exact for the
exactly solvable model of the $D$-component classical vector ``spins'' in
the limit $D\rightarrow \infty $. Nevertheless, for more realistic models
such as the classical Heisenberg model ($D=3$) and the Ising model ($D=1$),
it is very difficult to believe that the superparamagnetic relation holds
for all temperatures. Clearly, if the number ${\cal %
N}$ of atoms in the particle 
is large and the temperature is below $T_{c}$, then the argument of the
Langevin function in Eq. \negthinspace (\ref{spmrelation}) becomes large
already for so small fields that $M$ does not essentially deviate from its
zero-field value. Under these conditions Eq. \negthinspace (%
\ref{spmrelation}) should be a good approximation. On the other hand, for
smaller particles and near or above $T_{c},$ there should be deviations from
the simple behavior, the study of which is the purpose of this work.

The structure of the rest of this article is as follows. In
Sec.\negthinspace\ \ref{SecBasic} using the low-field expansion of $m$ and
general arguments we show that Eq. \negthinspace (\ref{spmrelation}) is {\it not exact} for any finite value of $D$ and nonzero
temperatures. In particular, in the high-temperature limit there is another
analytic form of Eq. \negthinspace (\ref{spmrelation}) with $B_{D}$
substituted by $B_{\infty }$. In Sec.\negthinspace\ \ref{SecSWT} we present
an explicit calculation of both $m(H,T)$ and $M(H,T)$ at low temperatures
with the help of a spin-wave theory which separates the global-rotation mode
and the ${\bf k}\neq {\bf 0}$ spin-wave modes. In Sec. \negthinspace \ref
{SecMC} we perform high-accuracy MC simulations for the Ising and classical
Heisenberg models in the box geometry to illustrate the superparamagnetic
behavior in a wide range of parameters.

\section{Basic Relations}

\label{SecBasic}

We use the classical spin-vector Hamiltonian

\begin{equation}
{\cal H}=-{\bf H}\sum_{i}{\bf s}_{i}-\frac{1}{2}\sum_{ij}J_{ij}{\bf s}%
_{i}\cdot {\bf s}_{j},\qquad |{\bf s}_{i}|=1,  \label{ham}
\end{equation}
where ${\bf s}$ is a $D$-component vector ($D=1$ for the Ising model and $%
D=3 $ for the Heisenberg model). For this Hamiltonian one can prove an
identity relating correlations functions and susceptibilities 
\begin{equation}
M^{2}=m^{2}+\frac{dm}{dx}+\frac{(D-1)m}{x},  \label{Mviam}
\end{equation}
where $x$ is given by Eq.\negthinspace\ (\ref{spmrelation}). On the
right-hand side of Eq.\negthinspace\ (\ref{Mviam}), the second and third
terms are contributions from the longitudinal and $D-1$ transverse
susceptibilities, respectively. This relation can be used to extract the
value of $M(H,T)$ from measurements of the induced magnetization $m$ and
susceptibilities. Let us demonstrate how it works at low fields, where the
expansion of $m$ can be written as 
\begin{equation}
m\cong \frac{a^{2}}{D}x-\frac{c^{4}}{D^{2}(D+2)}x^{3}.  \label{mloField}
\end{equation}
Applying Eq.\negthinspace\ (\ref{Mviam}) one readily obtains 
\begin{equation}
M\cong a+\frac{a^{4}-c^{4}}{2aD^{2}}x^{2}.  \label{MloField}
\end{equation}
At zero temperature the magnetic moment of the particle can be considered as
a rigid spin, thus in Eq.\negthinspace\ (\ref{mloField}) $a=c=1$ which
results in $M=1,$ independently of the field. At $T>0$ one has $a<1$ and $%
c<a,$ so that $M$ increases with the field. The coefficients $a$ and $c$ can
be calculated analytically at low and high temperatures [see, e.g.,
Eq.\negthinspace\ (\ref{acLT})]. Let us check now what happens if we try to
find $M$ from Eq.\negthinspace\ (\ref{spmrelation}) under the same
conditions. One can write 
\begin{equation}
m\cong \frac{M^{2}}{D}x-\frac{M^{4}}{D^{2}(D+2)}x^{3},\qquad M\cong
M_{0}+M_{2}x^{2},  \label{Bexpansion}
\end{equation}
and find the coefficients $M_{0}$ and $M_{2}$ from the condition that $m$
here coinsides with that of Eq.\negthinspace\ (\ref{mloField}). The result
is Eq.\negthinspace\ (\ref{MloField}) with $D^{2}\rightarrow D(D+2).$ This
is clearly a wrong result for any finite value of $D$ and nonzero
temperature. Only in the limit $T\rightarrow 0$ the coefficient $M_{2}$
vanishes and both approaches yield the same trivial result. Therefore, one
cannot use Eq.\negthinspace\ (\ref{spmrelation}) to take into account the
field variation of $M$ in the range where the argument of $B$ is of order
one or less. This formula can only be correct in the case of large particles
for which the change of $M$ in this field range is very small and $M$
actually changes for much larger fields where we already have $m\cong M.$

On the other hand, using these results one can find the correction to
Eq.\negthinspace\ (\ref{spmrelation}) at low fields. To this end, one can
write 
\begin{equation}
m=MB(Mx)+\delta ,  \label{spmrelationCorrLF}
\end{equation}
expand it for $x\ll 1$ using Eq.\negthinspace\ (\ref{MloField}) and equate
the result to Eq.\negthinspace\ (\ref{mloField}) to find $\delta $. The
result is 
\begin{equation}
\delta =-\frac{2(a^{4}-c^{4})}{D+2}\frac{x^{3}}{D^{3}}<0,  \label{deltaLF}
\end{equation}
that is, the Langevin function $B_{D}$ in Eq.\negthinspace\ (\ref
{spmrelation}) should be replaced by some function $F$ which goes {\it below}
$B_{D}$ at nonzero temperatures.

In the high-temperature limit one can find an explicit form of the
superparamagnetic relation which also differs from Eq. \negthinspace (\ref
{spmrelation}). Indeed, at high temperatures the exchange interaction can be
neglected and one has to solve a one-spin problem, which yields 
\begin{equation}
m=B(\xi ),\qquad M^{2}=m^{2}+\frac{1}{{\cal N}}\left( B^{\prime }(\xi )+(D-1)%
\frac{B(\xi )}{\xi }\right) ,  \label{HTspmrelation}
\end{equation}
where $\xi \equiv H/T.$ Using this relation, one can plot $\ m/M$ vs $%
x_{M}\equiv Mx$ and thus obtain the scaling function $F(x)$ which replaces $%
B_{D}(x)$ in Eq. \negthinspace (\ref{spmrelation}). For large particles, $%
{\cal N}\gg 1,$ in the relevant region $x_{M}\sim 1$ one has $\xi \ll 1$ and
the second of Eqs.\negthinspace\ (\ref{HTspmrelation}), with the use of $%
B^{\prime }(\xi )\cong B(\xi )/\xi \cong 1/D,$ simplifies to 
\begin{equation}
M^{2}=m^{2}+Dm/x.
\end{equation}
On the other hand, this relation holds in the large-$D$ limit for all
temperatures, particle sizes, and types of boundary conditions, and it can
be obtained from Eq.\negthinspace\ (\ref{Mviam}) by dropping the term $dm/dx$
and replacing $D-1\rightarrow D.$ Solving this equation for $m$ yields the
scaling function of the spherical model 
\begin{equation}
F(x)=B_{\infty }(x)=\frac{2x/D}{1+\sqrt{1+(2x/D)^{2}}},  \label{Binf}
\end{equation}
in Eq. \negthinspace (\ref{spmrelation}) which goes below $B_{D}(x)$ for any
finite $D.$

\section{Spin-wave theory for finite-size magnetic particles}

\label{SecSWT}

\subsection{General}

At low temperatures all spins in the particle are strongly correlated and
they form a ``giant spin'' ${\bf M}$ [see Eq.\negthinspace\ (\ref{MvecDef}%
)] which behaves superparamagnetically. In addition, there are internal
spin-wave excitations in the particle which are responsible at nonzero
temperatures for the fact that $M<1$ and for the field dependence of $M$. In
our case of three-dimensional particles, $d=3$, these excitations can be
described perturbatively in small deviations of individual spins ${\bf s}%
_{i} $ from the direction of ${\bf M.}$ To this end, it is convenient to
insert an additional integration over $d{\bf M}=M^{D-1} dMd{\bf n}$ in the
partition function, 
\begin{equation}
{\cal Z}=\int M^{D-1}dMd{\bf n}\prod_{i}d{\bf s}_{i}\delta \left( {\bf M-}%
\frac{1}{{\cal N}}\sum_{i}{\bf s}_{i}\right) {\rm e}^{-{\cal H}/T},
\label{PartFunc}
\end{equation}
and first integrate over the magnitude $M$ of the central spin [this
variable appears locally and it should not be confused with the intrinsic
magnetization $M$ defined by Eq.\negthinspace\ (\ref{Mdef})]. To do this,
one should reexpress the vector argument of the $\delta $-function in the
coordinate system specified by the direction of the central spin ${\bf n}$%
{\bf , }which yields 
\begin{eqnarray}
\delta \left( {\bf M-}\frac{1}{{\cal N}}\sum_{i}{\bf s}_{i}\right) &=&\delta
\left( M{\bf -}\frac{1}{{\cal N}}\sum_{i}({\bf n\cdot s}_{i})\right) 
\nonumber \\
&&\times \delta \left( \frac{1}{{\cal N}}\sum_{i}\left[ {\bf s}_{i}-{\bf n}(%
{\bf n\cdot s}_{i})\right] \right) .
\end{eqnarray}
Then after integration over $M$ one obtains 
\begin{equation}
{\cal Z}=\int d{\bf n}{\cal Z}_{{\bf n}},  \label{ZviaZn}
\end{equation}
where 
\begin{equation}
{\cal Z}_{{\bf n}}=\int \prod_{i}d{\bf s}_{i}\delta \left( \frac{1}{{\cal N}}%
\sum_{i}\left[ {\bf s}_{i}-{\bf n}({\bf n\cdot s}_{i})\right] \right) {\rm e}%
^{-{\cal H}_{{\rm eff}}/T}  \label{Zndef}
\end{equation}
and 
\begin{eqnarray}
{\cal H}_{{\rm eff}} &=&-({\bf n\cdot H})\sum_{i}({\bf n\cdot s}_{i})-\frac{1%
}{2}\sum_{ij}J_{ij}{\bf s}_{i}\cdot {\bf s}_{j}  \nonumber \\
&&-(D-1)T\ln \left[ \frac{1}{{\cal N}}\sum_{i}{\bf n\cdot s}_{i}\right] .
\label{Heff}
\end{eqnarray}
In Eq.\negthinspace\ (\ref{Zndef}), the $\delta $-function expresses the
obvious condition that the sum of all spins does not have a component
perpendicular to the central spin ${\bf M}$. This will lead to the absence
of the zero Fourier component of the transverse fluctuations of spins in the
particle. The corresponding global-rotation Goldstone mode (which is
troublesome in the standard spin-wave theory for finite systems) has been
transformed into the integration over the global variable ${\bf n}$ in
Eq.\negthinspace\ (\ref{ZviaZn}) in the present formalism. The condition
mentioned above was also used to transform the Zeeman term in
Eq.\negthinspace\ (\ref{Heff}). This describes now the spins ${\bf s}_{i}$
in a field in the direction ${\bf n}$ and with the strength ${\bf n\cdot H}$%
{\bf .} As we will see below, the last term in Eq.\negthinspace\ (\ref{Heff}%
) is nonessential in the leading approximation at low temperatures.

To calculate ${\cal Z}_{{\bf n}}$ at low temperatures, one can expand ${\cal %
H}_{{\rm eff}}$ up to the bilinear terms in the transverse spin components 
\begin{equation}
{\bf \Pi }_{i}\equiv {\bf s}_{i}-{\bf n}({\bf n\cdot s}_{i})
\label{Ptransdef}
\end{equation}
using 
\begin{equation}
{\bf n\cdot s}_{i}=\sqrt{1-{\bf \Pi }_{i}^{2}}\cong 1-{\bf \Pi }_{i}^{2}/2.
\end{equation}
This yields 
\begin{eqnarray}
{\cal H}_{{\rm eff}} &\cong &E_{0}-{\cal N}{\bf n\cdot H}+\frac 12 \sum_{ij}A_{ij}%
{\bf \Pi }_{i}\cdot {\bf \Pi }_{j},  \nonumber \\
A_{ij} &\equiv & \left[(D-1)T/{\cal N}+{\bf n\cdot H+}%
\sum_l J_{il} \right] \delta _{ij}-J_{ij},  \label{HamPtrans}
\end{eqnarray}
where $E_{0}=-(1/2)\sum_{ij}J_{ij}$ is the zero-field ground-state energy.  For the
lattice sites inside the particle or for the model with periodic boundary
conditions one has $z=2d,$ where $d$ is the spatial dimension; for the sites
on the boundaries $z_{i}<2d.$ Now ${\cal Z}_{{\bf n}}$ in Eq.\negthinspace\ (%
\ref{Zndef}) takes on the form 
\begin{eqnarray}
{\cal Z}_{{\bf n}} &\cong &{\rm \exp }\left( \frac{-E_{0}+{\cal N}{\bf %
n\cdot H}}{T}\right) {\cal N}^{D-1}\int_{-\infty }^{\infty
}\prod_{i}\prod_{\alpha =1}^{D-1}d\Pi _{i}^{\alpha }\,  \nonumber \\
&&\times \delta \left( \sum_{i}{\bf \Pi }_{i}\right) \exp \left( -\frac{1}{2T}%
\sum_{ij}A_{ij}{\bf \Pi }_{i}\cdot {\bf \Pi }_{j}\right) ,  \label{ZnMatr}
\end{eqnarray}
which after working out the Gaussian integral over $\Pi _{i}^{\alpha }$
yields 
\begin{equation}
{\cal Z}_{{\bf n}}\cong {\rm \exp }\left( \frac{-E_{0}+{\cal N}{\bf n\cdot H}%
}{T}\right) {\cal N}^{D-1}\left[ \frac{(2\pi T)^{{\cal N}-1}}{\det
A_{ij}^{\prime }}\right] ^{(D-1)/2},  \label{ZnDeterm}
\end{equation}
where the matrix $A_{ij}^{\prime }$ is obtained from $A_{ij}$ of
Eq.\negthinspace\ (\ref{HamPtrans}) by elimination of ${\bf \Pi }_{i}$ on
one of the ${\cal N}$ lattice sites using the condition $\sum_{i}{\bf \Pi }%
_{i}=0.$

Eq.\negthinspace\ (\ref{ZnDeterm}) is a general result which is valid for a particle of arbitrary shape and for different types of exchange interaction $J_{ij}$ and boundary conditions.
In the following subsection we will consider particles of cubic shape with the nearest-neighbour interactions and free and periodic boundary conditions (fbc and pbc).

\subsection{Free and periodic boundary conditions}

Let us express the matrix $A_{ij}$ through its eigenfunctions $f_{ki}$ as follows
\begin{equation}\label{Aijexp}
A_{ij} = \sum_k f_{ki}^* A_k f_{kj} ,
\end{equation}
where  $f_{ki}$ satisfy
\begin{equation}\label{fEq}
\sum_i f_{ki} A_{ij} = A_k f_{kj}
\end{equation}
and form an orthonormal and complete basis
\begin{equation}\label{fNorm}
 \sum_i f_{ki}^* f_{k'i} = \delta_{kk' }, \qquad  \sum_k f_{ki}^* f_{ki'} = \delta_{ii' }.
\end{equation}
In this basis, the sum over $ij$ in Eq.\negthinspace\ (\ref{ZnMatr}) can be rewritten as
\begin{equation}\label{sumA}
 \sum_{ij} A_{ij}{\bf \Pi }_{i}\cdot {\bf \Pi }_{j} =  \sum_k A_k {\bf \Pi }_k^* \cdot {\bf \Pi }_k ,
\end{equation}
where
\begin{equation}\label{PkDef}
{\bf \Pi }_k  \equiv \sum_i f_{ki} {\bf \Pi }_{i}.
\end{equation}

Now one can make the observation that in the set of eigenfunctions $f_{ki}$ there is one which 
is independent of $i$ and which can be conveniently ascribed to $k=0$, i.e., $f_0 = 1/{\cal N}$.
This follows since 
\begin{equation}\label{A0}
\sum_i A_{ij} = A_0 = (D-1)T/{\cal N}  + {\bf n\cdot H}
\end{equation}
is independent of $j$, $ A_0 $ being the zero-$k$ eigenvalue.
Now one can see that $\delta \left( \sum_{i}{\bf \Pi }_{i}\right)$ in 
Eq.\negthinspace\ (\ref{ZnMatr}) excludes integration over the zero
mode ${\bf \Pi }_0$ in the new representation. 
Fluctuations of the components ${\bf \Pi }_k $ with $k\neq 0$ yield
multiplicative contribitions to the partition function, so that one is
left with the integrals over $\Pi_k^\alpha$. 
If the eigenfunctions $f_{ki}$ are real, one obtains
\begin{equation}\label{Pintegr}
I_k^\alpha = \int_{-\infty}^\infty d  \Pi_k^\alpha \exp\left[-\frac {A_k}{2T} (\Pi_k^\alpha)^2 \right] =
\left( \frac {2\pi T } {A_k } \right)^{1/2}.
\end{equation}
If $f_{ki}$ are complex, one has to integrate independently over the
real and imaginary components $x_k^\alpha$ and $y_k^\alpha$ of
$\Pi_k^\alpha = x_k^\alpha + i y_k^\alpha$ which gives $2\pi T /A_k$. 
Complex eigenfunctions arise, however, only in the case of periodic
boundary conditions where, as we shall see, one has to take into
account only a half of the $k$ modes, which effectively restores the
result of  Eq.\negthinspace\ (\ref{Pintegr}). 
So let us consider for the moment only systems with real eigenfunctions. 
In this case, integrating over $({\cal N}-1)(D-1)$ modes (${\cal N}-1$
$k$-modes multiplied by $D-1$ transverse spin components) for  ${\cal
Z}_{{\bf n}}$  one obtains Eq.\negthinspace\ (\ref{ZnDeterm}) with  
\begin{equation}
\det A_{ij}^{\prime }=\prod_k {}^{^{\prime }} \frac{A_k} {\cal N},  \label{DetApbc}
\end{equation}
where the prime on the product means that the mode with $k=0$ is omitted. 

All the results above are still general. 
Now we will consider cubic-shaped particles with free and periodic boundary conditions.
In the fbc case the matrix $A_{ij}$ has the form
\begin{equation}\label{Afbc}
A_{ij} = A_0\delta_{ij} + \Delta _{ij}^{(x)} + \Delta _{ij}^{(y)} + \Delta _{ij}^{(z)},
\end{equation}
where $A_0$ is given by Eq.\negthinspace\ (\ref{A0}) and 
\begin{eqnarray}\label{Delfbc}
&&
 \Delta _{ij}^{(x)} = - J[ \delta_{i_x,j_x-1} + \delta_{i_x,j_x+1} - 
\delta_{i_x,j_x} ( 2 - \delta_{j_x,1} - \delta_{j_x,N} ) ] 
\nonumber\\
&&
\qquad\qquad {} \times \delta_{i_y,j_y} \delta_{i_z,j_z},
\end{eqnarray}
etc., $\delta_{i_x,j_x}$ are Kronecker symbols, and $i_x,j_x=1,\ldots,N$.
If $j_x-1$ or $j_x+1$ run out of the particle, the corresponding $
\delta_{i_x,j_x-1}$ or $ \delta_{i_x,j_x+1}$ should be omitted. 
One can see that $ \Delta _{ij}^{(x)} $ is a discrete Laplace operator
for the coordinate $x$, and the eigenvalue problem factorizes. 
The eigenfunctions are standing waves and they can be written in the form
\begin{equation}\label{SepVars}
f_{{\bf k}i} = f_{i_x,k_x} \times f_{i_y,k_y} \times f_{i_z,k_z},
\end{equation}
where for $\alpha = x,y,z$ 
\begin{eqnarray}\label{SepSol}
&&
f_{i_\alpha,k_\alpha} = \sqrt{ \frac 2 {(1+\delta_{k_\alpha,0})N}} \cos[(i_\alpha-1/2)k_\alpha],  
\nonumber\\
&&
k_\alpha = \pi n_\alpha/N, \qquad n_\alpha = 0, 1, \ldots , N-1.
\end{eqnarray}
For the eigenvalue $A_k$ one obtains
\begin{equation}\label{AkRes}
A_{\bf k} = A_0 + J_{\bf k}-J_0.
\end{equation}

In the case of periodic boundary conditions, one should drop the terms
$\delta_{j_x,1}$ and $\delta_{j_x,N}$ and identify  $i_x=N+1$ with
$i_x=1$   in Eq.\negthinspace\ (\ref{Delfbc}). 
The eigenfunctions can be conveniently taken in the form of plane
waves $e^{-i{\bf kr}_{i}}$ with the wave vectors quantized as
$k_{\alpha }=2\pi n_{\alpha }/N$, the eigenvalue $A_{\bf k}$ having
the same form as in the fbc case. 
That is, the pbc and fbc models differ only by the quantization of the
wave vector
\begin{equation}
k_{\alpha }=\left\{ 
\begin{array}{cc}
2\pi n_{\alpha }/N, & {\rm pbc} \\ 
\pi n_{\alpha }/N, & {\rm fbc}
\end{array}
\right. ,\qquad n_{\alpha }=0,1,...,N-1  \label{defkpbc}
\end{equation}
where $\alpha =x,y,z.$ 
This subtle difference is responsible for much stronger thermal
fluctuations in the fbc model due to surface effects, as we shall see
below.

\subsection{The partition function}

Collecting the formulae obtained above, one can write down the expession for 
${\cal Z}_{{\bf n}}$ in the form 
\begin{equation}
{\cal Z}_{{\bf n}}\cong \tilde A{\rm \exp }\left( {\bf n\cdot x}+{\cal N}\frac{D-1}{%
2}f(G_{{\bf n}})\right) ,  \label{Znf}
\end{equation}
where $\tilde A$ is given by the first line of Eq.\negthinspace\ (\ref{Aprimedef})
below,

\begin{equation}
{\bf x}\equiv {\cal N}{\bf H/}T  \label{xdef}
\end{equation}
is the reduced field, the function $f(G_{{\bf n}})$ is defined by 
\begin{equation}
f(G_{{\bf n}})\equiv \frac{1}{{\cal N}}\sum_{{\bf k}}{}^{^{\prime }}\ln 
\frac{G_{{\bf n}}}{1-G_{{\bf n}}\lambda _{{\bf k}}},  \label{fdef}
\end{equation}
with $\lambda _{{\bf k}}\equiv J_{{\bf k}}/J_{0}=(\cos k_{x}+\cos k_{y}+\cos
k_{z})/d,$ and 
\begin{equation}
G_{{\bf n}}\equiv \frac{1}{1+a_{{\bf n}}}\cong 1-a_{{\bf n}},\quad a_{{\bf n}%
}\equiv \frac{T}{{\cal N}J_{0}}({\bf n\cdot x+}D-1)\ll 1.  \label{Gdef}
\end{equation}
Note that the angular dependence of ${\cal Z}_{{\bf n}}$ is more complicated
than that for rigid spins because of the internal spin-wave modes described
by the last term in Eq.\negthinspace\ (\ref{Znf}). These SW modes have a gap
accounted for by the first two terms in the denominator of Eq.\negthinspace\
(\ref{DetApbc}) or the dimensionless parameter $a_{{\bf n}}$ in
Eq.\negthinspace\ (\ref{Gdef}). One contribition to the gap is due to the
finite size of the particle and the other is due to the magnetic field. The
latter depends on the orientation of the particle's magnetic moment ${\bf n}$
with respect to the field{\bf .}

\begin{figure}[t]
\unitlength1cm
\begin{picture}(11,5.7)
\centerline{\psfig{file=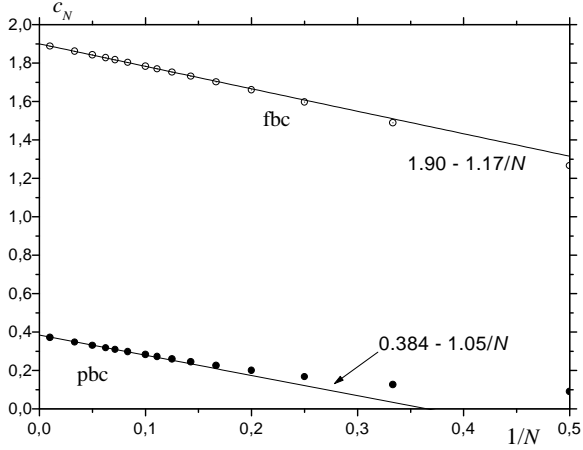,angle=-90,width=9cm}}
\end{picture}
\caption{ \label{fig-cn} 
Finite-size effect on $c_N\equiv f_P(0)$ [see Eq.\ (\protect\ref{Pasmall})]
for cubic systems with free and periodic boundary conditions.}
\end{figure}

The function $f(G_{{\bf n}})$ of Eq.\negthinspace\ (\ref{fdef}) can be
written as 
\begin{equation}
f(G_{{\bf n}})=f(1)-\int_{G_{{\bf n}}}^{1}du\frac{\widetilde{P}_{N}(u)}{u},
\label{fviaP}
\end{equation}
where $f(1)$ is a constant and 
\begin{equation}
\widetilde{P}_{N}(G)\equiv \frac{1}{{\cal N}}\sum_{{\bf k}}{}^{^{\prime }}%
\frac{1}{1-G\lambda _{{\bf k}}}  \label{Pdef}
\end{equation}
is the lattice Green function. Since at low temperatures the argument $G$ in
the expressions above is close to 1, it is convenient to write 
\begin{eqnarray}
\widetilde{P}_{N}(G) &=&\widetilde{P}_{N}(1)-\frac{1}{{\cal N}}\sum_{{\bf k}%
}{}^{^{\prime }}\frac{(1-G)\lambda _{{\bf k}}}{(1-G\lambda _{{\bf k}%
})(1-\lambda _{{\bf k}})}  \nonumber \\
\qquad \qquad \qquad  &=&W_{N}-N(1-G)f_{P}(1-G),  \label{Pasmall}
\end{eqnarray}
where $W_{N}\equiv \widetilde{P}_{N}(1).$ Here, if the linear size $N$ is
not large, one can replace $G\rightarrow 1$ in the argument of the function $%
f_{P}.$ For $N\gg 1$ the situation becomes more complicated since the wave
vectors ${\bf k}$ come closer to the origin and a singularity is formed. For
the system with free boundary conditions, the sum is dominated by $k\ll 1,$
so that $\lambda _{{\bf k}}\cong 1-k^{2}/(2d)$ and $f_{P}(y)$ has the form 
\begin{equation}
f_{P}(y)\cong \frac{(2d)^{2}}{\pi ^{4}}\sum_{n_{x},n_{y},n_{z}=0}^{\infty
}{}^{^{\prime }}\frac{1}{(y+n^{2})n^{2}}  \label{fPdef}
\end{equation}
with $n^{2}=n_{x}^{2}+n_{y}^{2}+n_{z}^{2}$ and 
\begin{equation}
y\equiv 2d(1-G)(N/\pi )^{2}.  \label{ydef}
\end{equation}
For $y\ll 1$ one can set $y=0$ which yields $f_{P}(y)\cong f_{P}(0)=c_{{\rm %
fbc}}\simeq 1.90,$ whereas for $y\gg 1$ one can replace summation by
integration and calculate the integral analytically. For the model with
periodic boundary conditions, there are different contributions from
different corners of the Brillouin zone in Eq.\negthinspace\ (\ref{Pasmall}%
), and one obtains a more cumbersome analogue of Eq.\negthinspace\ (\ref
{fPdef})$.$ In practice, it is easier to compute $f_{P}$ from its definition
in Eq.\negthinspace\ (\ref{Pasmall}). For three-dimensional cubic particles
the limiting cases are ($1-G\ll 1)$ 
\begin{equation}
\widetilde{P}_{N}(G)\cong W_{N}-\left\{ 
\begin{array}{ll}
c_{N}N(1-G), & y\ll 1 \\ 
c_{0}\sqrt{1-G}, & y\gg 1,\;
\end{array}
\right.   \label{Plims}
\end{equation}
where for l$\arg $e $N$ the value of $W_{N}$ approaches the Watson integral $%
W=1.51639$ according to \cite{kacgar01} 
\begin{equation}
\Delta _{N}\equiv \frac{W_{N}-W}{W}\cong \left\{ 
\begin{array}{ll}
\displaystyle
-\frac{0.90}{N}, & {\rm pbc} \\ 
\displaystyle
\frac{9\ln (1.17N)}{2\pi WN}, & {\rm fbc}
\end{array}
\right.   \label{DeltaN}
\end{equation}
(notice the positive sign for the fbc model). For the simple cubic lattice $%
c_{0}=(2/\pi )(3/2)^{3/2}$. The numerically obtained results for $c_{N}$
can, for $N\gg 1,$ be fitted as 
\begin{equation}
c_{N}\cong \left\{ 
\begin{array}{ll}
0.384-1.05/N, & {\rm pbc} \\ 
1.90-1.17/N, & {\rm fbc}
\end{array}
\right.   \label{cN}
\end{equation}
(see Fig. \negthinspace \ref{fig-cn}). The square-root term in
Eq.\negthinspace\ (\ref{Plims}) describes the spin-wave singularity in the
infinite system. From Eqs.\negthinspace\ (\ref{ydef}) and (\ref{Gdef}) it
follows that the crossover to the bulk spin-wave behavior occurs for the
values of the reduced field $x\gtrsim x_{S}\sim NJ_{0}/T$ which is much
larger than the value $x\sim x_{V}=1$ corresponding to the suppression of
the global rotation of the particle's magnetic moment. The actual crossover
fields, in notations of Ref.\negthinspace\ \cite{fispri85}, are given by 
\begin{equation}
H_{V}=\frac{T}{{\cal N}}\ll H_{S}=\frac{\pi ^{2}}{2d}\frac{J_{0}}{{\cal N}%
^{2/3}},  \label{Hcross}
\end{equation}
that is, they are widely separated from each other in our case $%
T/(NJ_{0})\ll 1.$ Thus the result for the function $f(G_{{\bf n}})$ of
Eq.\negthinspace\ (\ref{fviaP}), which with the help of Eq.\negthinspace\ (%
\ref{Pasmall}) can be written as 
\begin{equation}
f(G_{{\bf n}})\cong f(1)-a_{{\bf n}}W_{N}+N\int_{0}^{a_{{\bf n}%
}}da\,af_{P}(y_{a})  \label{fviaP1}
\end{equation}
with $y_{a}\cong 2da(N/\pi )^{2},$ can be simplified in different field
ranges.

%

For $H\ll H_{S}$ one can replace $f_{P}(y_{a})$ by $c_{N}$ to obtain, in
Eq.\negthinspace\ (\ref{Znf}), 
\begin{eqnarray}
{\cal N}\frac{D-1}{2}f(G_{{\bf n}}) &\cong &{\cal N}\frac{D-1}{2}f(1)-t({\bf %
n\cdot x}+D-1{\bf )}  \nonumber \\
&&\qquad {\bf +\alpha }({\bf n\cdot x}+D-1{\bf )}^{2},  \label{NDfn}
\end{eqnarray}
where 
\begin{equation}
t\equiv \frac{D-1}{2}\frac{W_{N}T}{J_{0}},\qquad \alpha \equiv \frac{%
(D-1)c_{N}}{4N^{2}}\left( \frac{T}{J_{0}}\right) ^{2}  \label{talphadef}
\end{equation}
are small parameters, $\alpha \ll t\ll 1.$ Since $\alpha x^{2}\sim
(H/H_{S})^{2}\ll 1,$ one can expand the partition function ${\cal Z}$ of
Eqs.\negthinspace\ (\ref{ZviaZn}) and (\ref{Znf}) with respect to the last
term of Eq. \negthinspace (\ref{NDfn}), which yields 
\begin{eqnarray}
{\cal Z} &\cong & \tilde A^{\prime }\int_{-1}^{1}du(1-u^{2})^{(D-3)/2}  \nonumber \\
&&\times \exp \left\{ \left[ 1-t+2\alpha (D-1)\right] ux\right\} \left[
1+\alpha (ux)^{2}\right]   \label{Zexpanded}
\end{eqnarray}
where 
\begin{eqnarray}
\tilde A^{\prime } &=&{\cal N}^{D-1}\left( 2\pi {\cal N}T/J_{0}\right) ^{(D-1)(%
{\cal N}-1)/2} e^{ -E_0/T}   \nonumber \\
&&\times e^{ (D-1)[ {\cal N}f(1)/2-t ] } S_{D-1}
\label{Aprimedef}
\end{eqnarray}
and $S_{D}=2\pi ^{D/2}/\Gamma (D/2)$ is the surface of the $D$-dimensional
unit sphere. In fact, we have left the term proportional to $\alpha ux$ in
Eq. \negthinspace (\ref{Zexpanded}) not expanded for the sake of
convenience. Integration in Eq. \negthinspace (\ref{Zexpanded}) results in 
\begin{equation}
{\cal Z}\cong {\cal Z}_{0}\{[1-t+2\alpha (D-1)]x\}+\alpha x^{2}\frac{d^{2}%
{\cal Z}_{0}}{dx^{2}},  \label{Zresult}
\end{equation}
where ${\cal Z}_{0}\{[1-t+2\alpha (D-1)]x\}$ is the partition function of
the rigid magnetic moment with the magnitude reduced by the factor $%
1-t+2\alpha (D-1)$.

\subsection{The superparamagnetic relation}

Using 
\begin{equation}
\frac{1}{{\cal Z}_{0}}\frac{d^{2}{\cal Z}_{0}}{dx^{2}}=B^{2}+B^{\prime }=1-%
\frac{(D-1)B}{x},  \label{Zsecder}
\end{equation}
where $B=B(x)$ is the Langevin function, for the induced magnetization $m$
one obtains 
\begin{eqnarray}
m &=&\frac{d\ln {\cal Z}}{dx}\cong \lbrack 1-t+2\alpha (D-1)]  \nonumber \\
&&\qquad \times B\{[1-t+2\alpha (D-1)]x\}  \nonumber \\
&&\qquad +\alpha \lbrack 2x-(D-1)(B+xB^{\prime })]  \nonumber \\
\qquad \qquad  &\cong &[1-t+\alpha (D-1)]  \nonumber \\
&&\times B\{[1-t+\alpha (D-1)]x\}+2\alpha x.  \label{minduced}
\end{eqnarray}
Expanding the expression for $m$ for $x\ll 1$ leads to Eq.\negthinspace\ (%
\ref{mloField}) with the explicit values of the parameters 
\begin{eqnarray}
a^{2} &=&[1-t+\alpha (D-1)]^{2}+2\alpha D,\qquad   \nonumber \\
c^{4} &=&[1-t+\alpha (D-1)]^{4}.  \label{acLT}
\end{eqnarray}
Note that in the region $x\gg 1,$ where a rigid magnetic moment would
saturate, $m$ continues to increase linearly as $m\cong 1-t+\alpha
(D-1)+2\alpha x.$ This is due to the field dependence of the intrinsic
magnetization $M$. The latter can be calculated from Eq.\negthinspace\ (\ref
{Mviam}) which leads to 
\begin{equation}
M\cong 1-t+\alpha (2D-1)+2\alpha xB(x).  \label{Mresult}
\end{equation}
This formula decribes a crossover from the quadratic field dependence of $M$
at low field, $x\ll 1,$ to the linear dependence at $x\gg 1.$

Now we are in a position to calculate the correction to Eq. \negthinspace (%
\ref{spmrelation}) at low temperatures and $x\sim 1.$ To this end, one can
write $m$ in the form of Eq.\negthinspace\ (\ref{spmrelationCorrLF}), expand
it with respect to $\alpha \ll 1$ and equate to the expanded form of Eq.
\negthinspace (\ref{spmrelation}). This gives 
\begin{equation}
\delta =\alpha \left[ 2x-(D+2xB)(B+xB^{\prime })\right] ,  \label{deltaLT}
\end{equation}
which has a negative value. In particular, for $x\ll 1$ one has $\delta
\cong -8\alpha x^{3}/[D^{2}(D+2)]$ [cf. Eq. \negthinspace (\ref{deltaLF})]$.$
It can be shown that $\delta \rightarrow 0$ in the large-$D$ limit. Since $%
\alpha $ defined by Eq. \negthinspace (\ref{talphadef}) contains $N^{2}$ in
the denominator, $\delta $ remains small even if $T\sim J_{0}.$ This is an
indication that Eq. \negthinspace (\ref{spmrelation}) is a very good
approximation for not extremely small systems in the whole range below $T_{c}
$. It can be shown that for $N\gg 1$ crossover to the high-temperature form
of Eq. \negthinspace (\ref{spmrelation}) specified by the function $%
B_{\infty }(x)$ of Eq. \negthinspace (\ref{Binf}) occurs in a close vicinity
of $T_{c}$.

At higher fields $H\sim H_{S}$ there is another crossover to the standard
spin-wave-theory result for $M.$ Here one has $x\gg 1,$ thus the integral in
Eq.\negthinspace\ (\ref{ZviaZn}) is dominated by ${\bf n\cdot x}\cong x.$
Replacing ${\bf n\cdot x}\rightarrow x$ in the last term of
Eq.\negthinspace\ (\ref{fviaP1}) one obtains 
\begin{eqnarray}
{\cal Z} &\cong &{\cal Z}_{0}[(1-t)x]\exp \left[ \frac{D-1}{2}{\cal N}%
^{4/3}\int_{0}^{a_{x}}da\,af_{P}(y_{a})\right] \qquad  \nonumber \\
a_{x} &\cong &xT/({\cal N}J_{0})=H/J_{0},  \label{Zhifield}
\end{eqnarray}
which yields 
\begin{equation}
m\cong M\cong 1-\frac{D-1}{2}\frac{T}{J_{0}}\left[ W_{N}-\frac{H}{J_{0}}%
Nf_{P}\left( \frac{H}{H_{S}}\right) \right] ,  \label{mMfifield}
\end{equation}
where the function $f_{P}(y)$ is defined by Eq.\negthinspace\ (\ref{Pasmall}%
) and $H_{S}$ is defined by Eq.\negthinspace\ (\ref{Hcross}).

Let us now write down the explicit forms of the field dependence of the
intrinsic magnetization $M$ in the three different field regions 
\begin{equation}
M\cong 1-t+\left\{ 
\begin{array}{ll}
\displaystyle
\frac{D-1}{2D}c_{N}\left( \frac{HN^{2}}{J_{0}}\right) ^{2}, & H\ll H_{V} \\ 
\displaystyle
\frac{D-1}{2}c_{N}\frac{NHT}{J_{0}^{2}}, & H_{V}\ll H\ll H_{S} \\ 
\displaystyle
\frac{D-1}{2}c_{0}\frac{T}{J_{0}}\left( \frac{H}{J_{0}}\right) ^{1/2}, & 
H_{S}\ll H\ll J_{0}.
\end{array}
\right.   \label{Mfinalres}
\end{equation}
Here $t\ll 1$ is defined by Eq.\negthinspace\ (\ref{talphadef}). In the
second and third field ranges, the particle's magnetic moment is fully
oriented by the field, thus $m\cong M,$ the spin-wave gap in
Eq.\negthinspace\ (\ref{DetApbc}) has the value $H$, as in the bulk, and the
field dependence of both magnetizations follows that of the function $\widetilde{P%
}_{N}(G)$ of Eqs.\negthinspace\ (\ref{Pasmall}) or (\ref{Plims}) with $%
1-G\cong H/J_{0}$ [see Eq.\negthinspace\ (\ref{Gdef})]. The region $H\ll
H_{V}$ in Eq.\negthinspace\ (\ref{Mfinalres}) is less trivial. Here the gap
in Eq.\negthinspace\ (\ref{DetApbc}) is ${\bf n\cdot H}$ and depends on the
orientation of the particle's magnetic moment which is not yet completely
aligned by the field. Effectively one has in this region ${\bf n\cdot H\sim }%
H^{2},$ which leads to a quadratic field dependence of $M$. In fact, such a
dependence at smallest fields already follows from general principles, see
Sec. \ref{SecBasic}, and is pertinent to the Ising model as well.

To conclude this subsection, we introduce the orientation-dependent
``macroscopic'' particle's magnetization ${\bf M}_{{\bf n}}$ according to 
\begin{equation}
{\bf M}_{{\bf n}}\equiv \frac{\partial \ln {\cal Z}_{{\bf n}}}{\partial {\bf %
x}},  \label{Mndef}
\end{equation}
where ${\cal Z}_{{\bf n}}$ and ${\bf x}$ are defined by Eqs.\negthinspace\ (%
\ref{Zndef}) and (\ref{xdef}), respectively. 
Using this definition, for the
induced magnetization ${\bf m\equiv \partial }{\ln {\cal Z}}/\partial {\bf x}
$ one can write 
\begin{equation}
{\bf m}=\frac{\int d{\bf n\,M}_{{\bf n}}{\cal Z}_{{\bf n}}}{\int d{\bf n}%
{\cal Z}_{{\bf n}}}.  \label{mviaMn}
\end{equation}
${\bf M}_{{\bf n}}$ can be interpreted as ${\bf M}$ of Eq.\negthinspace\ (%
\ref{Mdef}) with the spin-wave modes integrated out. From Eq.\negthinspace\ (%
\ref{Znf}) one obtains 
\begin{equation}
{\bf M}_{{\bf n}}=\left[ 1-\frac{D-1}{2}\frac{T}{J_{0}}\frac{\widetilde{P}%
_{N}(G_{{\bf n}})}{G_{{\bf n}}}\right] {\bf n,}  \label{Mnresult}
\end{equation}
which for $H\ll H_{S}$ can be written as 
\begin{equation}
{\bf M}_{{\bf n}}\cong \left[ 1-t+2\alpha ({\bf n\cdot x}+D-1)\right] {\bf n.%
}  \label{Mnresult1}
\end{equation}
The magnitude of the particle's magnetization, $M_{{\bf n}}\equiv |{\bf M}_{%
{\bf n}}|,$ depends on its orientation due to spin-wave effects. It attains
its maximal value $1-t+2\alpha (x+D-1)$ if the particle's magnetization is
directed along the field ${\bf H}$ and its minimal value $1-t+2\alpha
(-x+D-1)$ in the thermodynamically unfavorable state with magnetization
against the field. It should be stressed that in order to obtain the
explicit result for the induced magnetization, Eq.\negthinspace\ (\ref
{minduced}), from Eq.\negthinspace\ (\ref{mviaMn}), one should know ${\cal Z}%
_{{\bf n}},$ so its calculation in the main part of this section is
unavoidable. On the other hand, for the intrinsic magnetization $M$ it is
sufficient to replace ${\cal Z}_{{\bf n}}\Longrightarrow \exp ({\bf n\cdot x)%
}$ and use

\begin{equation}
M\cong \frac{\int d{\bf n\,}M_{{\bf n}}\exp ({\bf n\cdot x)}}{\int d{\bf n}%
\exp ({\bf n\cdot x)}}  \label{MviaMorient}
\end{equation}
which readily yields Eq.\negthinspace\ (\ref{Mresult}) up to a
field-independent term ($\alpha \ll t)$.

\subsection{Local magnetization}

The formalism developed above can be applied to study inhomogeneities
in the particle's magnetization arising as a consequence of free boundaries. 
Since in zero field the standardly defined local {\em induced}
magnetization ${\bf m}_i \equiv \langle {\bf s}_i \rangle $ of a
finite-size particle vanishes,  one has to introduce  local {\em
intrinsic} magnetization 
\begin{equation}\label{Mi}
M_{i}=\frac{1}{M}\left\langle {\bf s}_{i}\cdot \frac{1}{\cal N}
\sum_{j} {\bf s}_{j}\right\rangle.
\end{equation}
One can check the identity $(1/{\cal N})\sum_{i}M_{i}=M$ showing the
self-consistency of the definition given above.
Adding the expression within the brackets to the integrand of
Eq.\negthinspace\ (\ref{PartFunc}) and repeating all operations, one
arrives at the final low-temperature result 
\begin{equation}\label{MiRes}
M_{i} \cong 1 - \frac{D-1 } 2 \frac T {\cal Z} \int d{\bf n}{\cal Z}_{{\bf n}}
 \sum_k {}^{'} \frac {|f_{ki}|^2}{A_k},
\end{equation}
where $A_k$ and $f_{ki}$ are eigenvalues and eigenfunctions of the
linear problem, see Eq.\negthinspace\ (\ref{fEq}). 
The latter contain the information about inhomogeneities in the system.
For periodic boundary conditions, one has $f_{ki}=e^{-i{\bf
kr}_{i}}/\sqrt{N}$, so that $|f_{ki}|=\sqrt{1/N}$ and there are no
inhomogeneities. 
Since the parameter $t$ defined by Eq.\negthinspace\ (\ref{talphadef})
is small, one can expand Eq.\negthinspace\ (\ref{MiRes}) to obtain, to
the lowest order at low temperatures, 
\begin{equation}\label{MiResFin}
M_{i} \cong 1 - \frac{(D-1) T} 2  \sum_k {}^{'} \frac {|f_{ki}|^2}{A_k}.
\end{equation}
Here one can check again $(1/{\cal N})\sum_{i}M_{i}=M$, where $M\cong
1-t$, according to  Eq.\negthinspace\ (\ref{Mresult}). 
For cubic particles with free boundary conditions, one has
$f_{ki}\approx \sqrt{2/N}$ at the boundary according to
Eq.\negthinspace\ (\ref{SepSol}), which is larger than the
bulk-averaged value. 
The biggest effect of the surface is naturally attained at the corners
of the cube where $M_i\approx 1-8t$.

\section{MC simulations and results}

\label{SecMC}

The classical Monte Carlo (MC) method based on the Metropolis algorithm is
now a standard technique (see, e.g., Ref. \cite{binhee92} for details). The
general idea is to simulate the statististics of a magnetic system by
generating a Markov chain of spin configurations and taking an average over
the latter. Each step of this chain (a MC step) is a stochastic transition
of the system from one state to another, subjected to the condition of the
detailed balance. Usually a MC step consists in generating a new trial
orientation of a spin vector on a lattice site $i$ and calculating the
ensuing energy change of the system $\Delta E.$ The trial confuguration is
accepted as a new configuration if 
\begin{equation}
\exp (-\Delta E/T)\geq {\cal R}(0,1),  \label{Metropolis}
\end{equation}
where ${\cal R}(0,1)$ is a random number in the interval $(0,1)$, otherwise
the old configuration is kept. As follows from Eq.\negthinspace\ (\ref
{Metropolis}), for $\Delta E\leq 0$ the trial orientation is accepted with a
probability 1. The trial orientation can be a completely random orientation,
or a random orientation in the vicinity of the initial orientation of the
spin ${\bf s}_{i},$ which is more appropriate at low temperatures. For the
Ising model, the trial orientation is generated by a flip of $s_{i}$ with a
probability 1/2. The MC steps are performed sequentially or randomly for all
lattice sites. This conventional version of the MC method is not efficient
for systems of finite size at low temperatures and small fields, if one is
interested in the induced magnetization $m$. The Boltzmann distribution over
the directions of the particle's magnetic moment ${\bf M}$ of Eq.
\negthinspace (\ref{MvecDef}) is achieved by rotations of ${\bf M}$ itself
rather than by rotations of individual spins ${\bf s}_{i}.$ Indeed, each
spin ${\bf s}_{i}$ is acted upon by the strong exchange field ${\bf H}%
_{E,i}=\sum_{j}J_{ij}{\bf s}_{j}\sim J_{0},$ and in the typical case $H\ll
J_{0}$ all trial configurations with the direction of ${\bf s}_{i}$
significantly differing from that of its neighbors are rejected with a
probability close to 1. Thus in the standard MC procedure directions of
individual spins can only change little by little, and the resulting change
of ${\bf M}$ is extremely slow. For the Ising model the situation is even
worse since the spin geometry is discrete and there are no small changes of
spin directions, whereas a flip of a single spin against the exchange field
has an exponentially small probability. Hence if one starts in zero field
with the configuration of all spins up or all spins down, the magnetization $%
m$ will practically never relax to zero. This drawback can be remedied by
augmenting the procedure by a global rotation (GL) of the particle's spins
to a new trial direction of ${\bf M}$ and calculating the energy change.
That is, before turning single spins on all lattice sites, one computes $%
{\bf M,}$ generates its new orientation ${\bf M}^{\prime }$ and obtains the
energy difference 
\begin{equation}
\Delta E=-{\cal N}{\bf H\cdot (M}^{\prime }-{\bf M).}  \label{GlobRotDeltaE}
\end{equation}
If the new orientation is accepted according to Eq.\negthinspace\ (\ref
{Metropolis}), one turns all spins ${\bf s}_{i}$ by an appropriate angle and
proceeds with the standard Metropolis method recapitulated above. In small
fields ($x\equiv {\cal N}H/T\lesssim 1$) relaxation of the induced
magnetization $m$ becomes much slower than that of the intrinsic
magnetization $M,$ and one needs much more MC steps to find the former than
the latter with the same precision. If in the procedure each global rotation
is coupled with subsequent turning of single spins on all lattice sites $i$,
making enough global rotations to achieve a required precision for $m$ costs
much more computer time for larger particle sizes. Thus a natural idea is to
make many global rotations and gather the data for $m$ after each GL before
proceeding to the conventional (single-spin) part of the Metropolis
algorithm. This improved method is especially fast for the isotropic
Heisenberg or Ising models where the energy change is given by
Eq.\negthinspace\ (\ref{GlobRotDeltaE}) since, after ${\bf M}$ has been
initially computed, each of its subsequent rotations and calculations of $%
\Delta E$ requires $O(1)$ operations. In contrast, for systems with
anisotropy one has to perform a sum over all lattice sites for each
orientation of ${\bf M,}$ i.e., to make $O({\cal N})$ operations${\bf
.}$

Finally, we mention that for the Heisenberg model $5^3$ the running
time of our programme with global rotations on a Pentium III/933 MHz
is 160mn, for a precision of $10^{-4} - 10^{-5}$ on the magnetisation.

\begin{figure}[t]
\unitlength1cm
\begin{picture}(11,5.7)
\centerline{\psfig{file=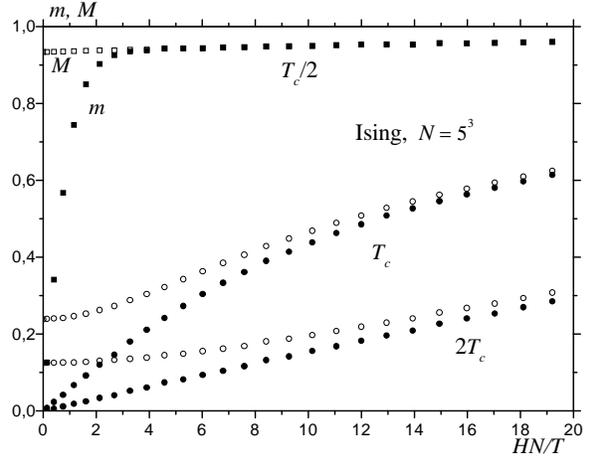,angle=-90,width=9cm}}
\end{picture}
\caption{ \label{fig-i} 
Field dependence of the intrinsic magnetization $M$ and the induced magnetization $m$ of the Ising model on the sc lattice with fbc for different temperatures.
}
\end{figure}
%

\begin{figure}[t]
\unitlength1cm
\begin{picture}(11,5.7)
\centerline{\psfig{file=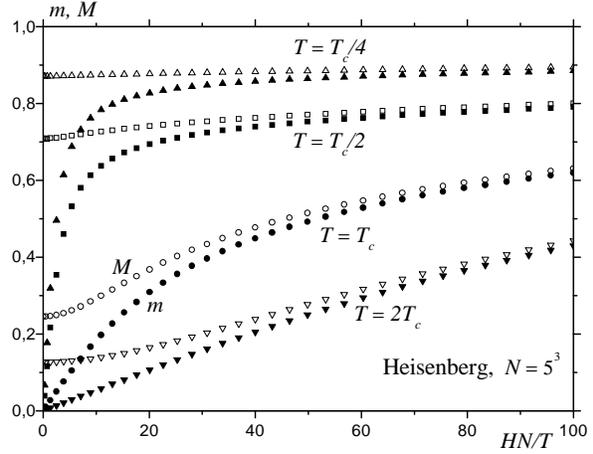,angle=-90,width=9cm}}
\end{picture}
\caption{ \label{fig-h} 
Field dependence of the intrinsic magnetization $M$ and the induced magnetization $m$ of the Heisenberg model on the sc lattice with fbc for different temperatures.
}
\end{figure}
%

Figs. \negthinspace \ref{fig-i} and \ref{fig-h} show the results of our MC
simulations for the Ising and Heisenberg models on a cubic lattice with size 
${\cal N}=5^{3}$ and free boundary conditions. The intrinsic magnetization $M
$ and induced magnetization $m$ are plotted vs the scaled field $x\equiv 
{\cal N}H/T$ for different temperatures. We used the bulk Curie temperatures 
$T_{c}=\theta _{c}T_{c}^{{\rm MFA}},$ where $T_{c}^{{\rm MFA}}=J_{0}/D$ is
the mean-field Curie temperature and $\theta _{c}$ is 0.751 for the Ising
model and 0.722 for the Heisenberg model. One can see that the particle's
magnetic moment is aligned and thus $m\sim M$ for $x\gtrsim 1,$ if $T\ll
T_{c}.$ At $T\gg T_{c}$ the field alignes individual spins and this requires 
$H\gtrsim T,$ i.e., $x\gtrsim {\cal N}.$ The quadratic dependence of $M(H)$
at small fields, which is phenomenologically described by Eq.\negthinspace\ (%
\ref{MloField}), manifests itself strongly at elevated temperatures. At low
temperatures this dependence is much more difficult to see on the graph
because the field-dependent part of $M,$ which for the Heisenberg model is
given by Eq.\negthinspace\ (\ref{Mresult}), is for $x\sim 1$ proportional to 
$\alpha $ of Eq.\negthinspace\ (\ref{talphadef}), which is very small. For
the Ising model there is practically no field dependence of $M$ at low
temperatures since $M$ is very close to 1. The weak linear field dependence
of $M$ for the Heisenberg model which is visible on Fig. \negthinspace \ref
{fig-i} at $T=T_{c}/4$ will be quantitatively explained below.

\begin{figure}[t]
\unitlength1cm
\begin{picture}(11,5.7)
\centerline{\psfig{file=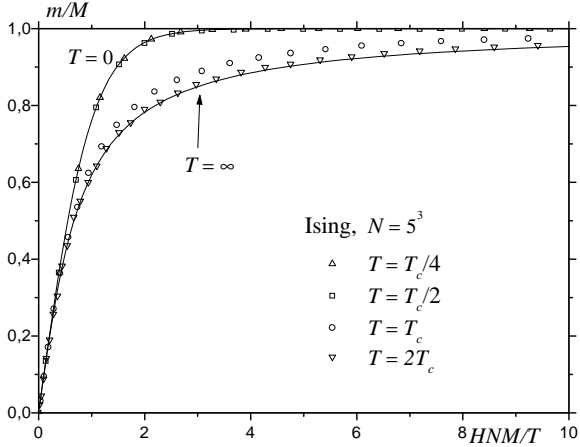,angle=-90,width=9cm}}
\end{picture}
\caption{ \label{fig-is} 
Scaled graph for the induced magnetization $m$ of the Ising model on the sc lattice with fbc for different temperatures. Theoretical curves at low temperatures, $B_1(x)=\tanh x$, and at high temperatures, $B_\infty(x)$, are shown by solid lines.
}
\end{figure}
%

\begin{figure}[t]
\unitlength1cm
\begin{picture}(11,5.7)
\centerline{\psfig{file=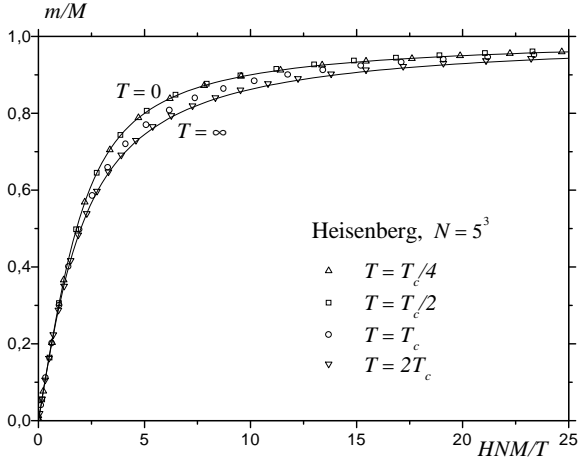,angle=-90,width=9cm}}
\end{picture}
\caption{ \label{fig-hs} 
Scaled graph for the induced magnetization $m$ of the Heisenberg model on the sc lattice with fbc for different temperatures. Theoretical curves at low temperatures, $B_3(x)=\coth x -1/x$, and at high temperatures, $B_\infty(x)$, are shown by solid lines.
}
\end{figure}
%

Figs. \negthinspace \ref{fig-is} and \ref{fig-hs} show that the
superparamagnetic relation of Eq.\negthinspace\ (\ref{spmrelation}) with the
Langevin function $B_{D}(x)$ in place of $F(x)$ is a very good approximation
everywhere below $T_{c},$ for both Ising and Heisenberg models. On the other
hand, above $T_{c}$ Eq.\negthinspace\ (\ref{spmrelation}) with the function $%
B_{\infty }(x)$ of Eq.\negthinspace\ (\ref{Binf}) is obeyed. The difference
between these limiting expressions decreases with increasing the number $D$
of spin components and disappears in the spherical limit ($D\rightarrow
\infty $).

\begin{figure}[t]
\unitlength1cm
\begin{picture}(11,5.7)
\centerline{\psfig{file=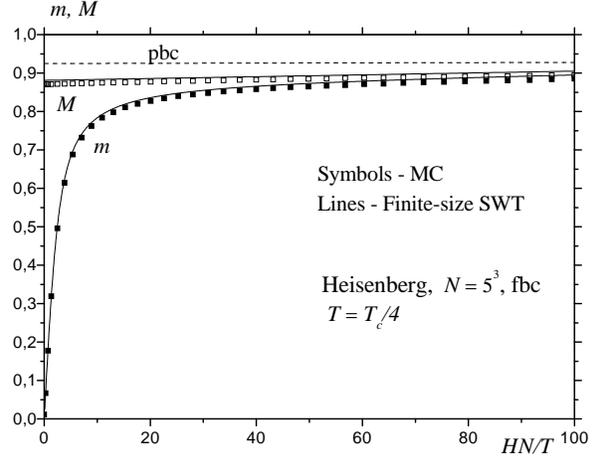,angle=-90,width=9cm}}
\end{picture}
\caption{Comparison of the theoretical and MC results for the field
dependences of the magnetizations $M$ and $m$ for the Heisenberg model at $T=T_c/4$. }
\label{fig-hth}
\end{figure}
%

In Fig.\negthinspace\ \ref{fig-hth} we compare theoretical predictions of
Sec.\negthinspace\ \ref{SecSWT} for the Heisenberg model at $T=T_{c}/4$ with
our MC results. For our small size ${\cal N}=5^{3}$ the square-root field
dependence of the magnetization [the third line of Eq.\negthinspace\ (\ref
{Mfinalres})] does not arise and finite-size corrections are very important.
For $M$ one should use Eq.\negthinspace\ (\ref{Mresult}), where $t$ and $%
\alpha $ are given by Eq.\negthinspace\ (\ref{talphadef}) with the
numerically exact values $W_{N}=1.99$ and $c_{N}=1.66$ for the fbc model
[cf. Eqs.\negthinspace\ (\ref{DeltaN}) and (\ref{cN})]. This yields $t\simeq
0.119$ and $\alpha \simeq 1.20\times 10^{-4}.$ The corresponding theoretical
dependence $M(H)$ is practically a straight line which goes slightly above
the MC points. This small discrepancy can be explained by the fact that the
applicability criterion of our analytical method, $t\ll 1,$ is not strongly
satisfied. For comparison we also plotted the theoretical $M(H)$ for the
unrealistic model with periodic boundary conditions. Here one has $W_{N}=1.25
$ and $c_{N}=0.20,$ thus $t\simeq 0.075$ and $\alpha \simeq 1.45\times
10^{-5},$ so $M(H)$ goes noticeably higher and with a much smaller slope.
The quadratic field dependence of $M$ in the region $x\lesssim 1$ is not
seen at this low temperature since the value of $\alpha $ is very small and
thus much more accurate MC simulations are needed. We have not performed
these simulations because the corresponding effects are very small. We also
plotted in Fig.\negthinspace\ \ref{fig-hth} the field dependence of $m$
given by Eq.\negthinspace\ (\ref{minduced}) in comparison with our MC data.
The agreement is reasonably good for $m$ as well.

\section{Discussion}

We have performed analytical and numerical investigation of the magnetic
field dependence of the intrinsic magnetization $M$ and induced
magnetization $m$ of the Ising and isotropic classical Heisenberg models on
cubic lattices of finite size. For the latter, we obtained explicit
analytical results for both $M(H,T)$ and $m(H,T)$ at low temperatures with
the help of a spin-wave theory singling out the global-rotation mode. These
results are in accord with our MC simulation data.

We investigated the validity of the superparamagnetic relation $%
m(H,T)=M(H,T)B_{D}[M(H,T){\cal N}H/T],$ where $B_{D}(x)$ is the Langevin
function and $D$ is the number of spin components. Both general arguments of
Sec.\negthinspace\ \ref{SecBasic} and explicit low-temperature results for
the Heisenberg model show that this is {\it not} an exact relation for any
finite $D$. Nevertheless, it is an extremely good approximation in the whole
range below $T_{c}$ for not too small particles, since, for the Heisenberg
model, its error behaves as $[T/(J_{0}N)]^{2},$ where $N$ is the linear
particle size. For $N\gg 1,$ a crossover to the high-temperature form of the
relation above, which utilizes the Langevin function of the spherical model $%
B_{\infty }(x),$ occurs in a close vicinity of $T_{c}$. The difference
between the low- and high-temperature forms of the superparamagnetic
relation decreases with $D$ and disappears in the spherical limit, rendering
this relation exact \cite{kacgar01,kacnogtrogar00}.

\bigskip

D. A. Garanin is endebted to the CNRS and Laboratoire de Magn\'{e}tisme et
d'Optique for the warm hospitability extended to him during his stay in
Versailles in October-December 2000.


\end{document}